\documentclass[aps,onecolumn,groupedaddress,showpacs,nofootinbib]{revtex4}
\usepackage{epsfig}
\def\er#1#2{\relax\ifmmode{}^{+#1}_{-#2}\else$^{+#1}_{-#2}$\fi}
\newcommand{\be}{\begin{equation}}
\newcommand{\bea}{\begin{eqnarray}}
\newcommand{\ee}{\end{equation}}
\newcommand{\eea}{\end{eqnarray}}

\newcommand{\krig}[1]{\stackrel{\circ}{#1}}
\newcommand{\bd}{\bar d}
\newcommand{\eb}{\bar e}
\newcommand{\ef}{\bar f}
\newcommand{\od}{{\cal O}}
\def\({\Big(}
\def\){\Big)}

\def\slashchar#1{\setbox0=\hbox{$#1$}
   \dimen0=\wd0 \setbox1=\hbox{/} \dimen1=\wd1
   \ifdim\dimen0>\dimen1 \rlap{\hbox to \dimen0{\hfil/\hfil}} #1
   \else  \rlap{\hbox to \dimen1{\hfil$#1$\hfil}} / \fi}

\begin{document}
\def\footnoterule{\kern-3pt \hrule width\hsize \kern3pt}

\title{
Improved Unitarized Heavy Baryon Chiral Perturbation Theory
for $\pi N $ Scattering to fourth order
}
\author{A. G\'omez Nicola$^1$}\email{email:gomez@fis.ucm.es}
\author{J. Nieves$^2$}\email{email:jmnieves@ugr.es}
\author{J.R. Pel\'aez$^1$}\email{email:jrpelaez@fis.ucm.es},
\author{E. Ruiz Arriola$^2$}\email{email:earriola@ugr.es}
\affiliation{$^1$Departamento de F\'{\i}sica Te\'orica. 
Universidad Complutense. 28040 Madrid, Spain.}
\affiliation{$^2$Departamento de F\'{\i}sica Moderna, Universidad de
Granada, E-18071 Granada, Spain}

\thispagestyle{empty}

\begin{abstract}
We extend our previous analysis of the unitarized pion-nucleon
 scattering amplitude including up to fourth order terms in Heavy
 Baryon Chiral Perturbation Theory. We pay special attention to the
 stability of the generated $\Delta(1232)$ resonance, the convergence
 problems and the power counting of the chiral parameters.
\end{abstract}


\maketitle

\section{Introduction}

Unitarization methods have been widely and successfully employed in the
recent past to enlarge the applicability region
of Chiral Perturbation Theory (ChPT) expansions, both in the meson-meson sector as well as in the meson
baryon sector and  to describe the lightest resonances without
including them explicitly as degrees of freedom. Two important constraints are required: exact unitarity 
and compliance with Chiral
Perturbation Theory at a given order of the expansion. 
In practice this approach provides a remarkable 
description of data in the scattering region. In the case
of $\pi N $ scattering in the elastic region, the subject of this
paper, a thorough partial wave analysis exists~\cite{AS95} (see also
the recent update~\cite{Arndt:2003if}).  For such a system, pions and
nucleons are treated as explicit degrees of freedom and a consistent
counting becomes possible if nucleons are treated as heavy particles
but in a covariant framework \cite{IW89}, yielding the so called
Heavy-Baryon Chiral Perturbation Theory (HBChPT)
\cite{JM91,BK92,BKM95}. In this counting, the expansion for the
scattering amplitude is done as a series of $ e^N /( F^{2l} M^{N+1-2l} )
$ terms , with $l=1, \dots , [(N+1)/2] $, $M$ the baryon mass, and $F$ the pion decay constant. The
quantity $e$ is a generic parameter with dimensions of energy
constructed in terms of the pseudoscalar momenta and the velocity
$v^\mu$ ($ v^2 =1$ ) and off-shellness $ k $ of the baryons defined
through the equation $p_B=\krig{M}v +k $, with $p_B$ the baryon four
momentum and $\krig{M}$ the baryon mass at leading order in the
expansion. After the relevant effective Lagrangian was written
down~\cite{EM96}, and the issue of wave function renormalization was
studied~\cite{Ecker:1997dn} standard HBChPT calculations to
second~\cite{BKM97} third~\cite{Mo98,FMS98} and fourth~\cite{feme4}
order have become available. The unitarization of these amplitudes of
$\pi N$ scattering in the elastic region has followed closely these
developments, particularly the third order
calculation~\cite{Mo98,FMS98}. This is the lowest order which
generates a perturbative unitarity correction of the amplitude. The
unitarization was carried out either using the standard Inverse
Amplitude Method~\cite{GP99} (IAM) or its improved version~\cite{GNPR00}~\footnote{For an alternative scheme based
on the Bethe-Salpeter equation applied to the P33 channel see
Ref.~\cite{Nieves:2000km}}. By successful we mean the possibility of
describing the data in the resonance region with parameters of natural
size. The purpose of the present paper is to extend the study
initiated in Ref.~\cite{GNPR00} and to analyze specifically the
qualitative and quantitative new effects generated by the fourth order
contribution calculated in Ref.~\cite{feme4} in our unitarization
scheme. 

Let us then specify the scope and motivations of our work:
First, our scheme is based on two fundamental ideas: demanding exact
unitarity and considering the $F^{-2}$ HBChPT expansion independent
from and converging faster than the $M^{-1}$ one. In \cite{GNPR00}
we showed that this allows to generate the $\Delta(1232)$ as well as
to fit the remaining $S$ and $P$ wave channels with natural values for
the low-energy constants (LEC) unlike for instance the IAM \cite{GP99}. Our method was implemented in
\cite{GNPR00} with the first contribution of order $F^{-4}$ only,
coming from the third order amplitude. Including the fourth order will
allow us to check the convergence of our method by considering, for instance, the
$\od(F^{-4}M^{-1})$, to be included in the third order $F^{-4}$ term.

Second, there is an interesting issue that we did not account for in
\cite{GNPR00} which has to do with the separation of the dimensionful
third and fourth order LEC into two pieces contributing to the orders
$F^{-2}$ and $F^{-4}$. As we will see below, taking into account this
effect may change considerably our description of the partial waves.
The reason why we did not consider it in \cite{GNPR00} is that we used
the amplitudes in \cite{Mo98}, which provide an specific separation
that turns out to be very natural, as we will see below~\footnote{A similar
situation has appeared already in the NNLO unitary analysis of
$\pi\pi$ scattering~\cite{Nieves:2001de}.}.  
 
Third, comparing the perturbative results to order three \cite{FMS98}
and four \cite{feme4}, one observes that in order to achieve a
reasonable convergence, the fourth order constants become of unnatural
size and, furthermore, their particular values are often incompatible
from one fit to another. This is a signal of the bad convergence of
the HBChPT series and could influence also the convergence of our
unitarized formula.

Fourth, unitarization methods are rarely applied
beyond the leading order in the imaginary part of the amplitudes \cite{Nieves:2001de,Dobado:2001rv}.
The study of the fourth order of $\pi N$ system
within HBChPT provides an opportunity to learn about the unitarization
approach beyond this lowest order.

For comparison purposes with previous
works~\cite{BKM97,Mo98,FMS98,feme4,GP99,GNPR00} we will
take the partial wave analysis performed in Ref.~\cite{AS95}.  The
recent update~\cite{Arndt:2003if} does not bring significant changes
to our discussion.

\section{The unitarized amplitude}

In order to have a neat separate expansion of the partial waves in
powers of $M^{-1}$ and $F^{-2}$, we need to re-expand the amplitudes
in \cite{feme4}, as it was already done to third
order in \cite{GNPR00} with those in \cite{Mo98}. Then, following the notation in
\cite{GNPR00}, we have, to fourth order, for any partial wave
\begin{eqnarray}
f_{l \, \pm}^{(1) \, \pm} &=&
{m\over F^2} t_{l \, \pm}^{(1,1) \, \pm} \({\omega \over m}\) \nonumber\\
f_{l \, \pm}^{(2) \, \pm} &=&
{m^2 \over F^2 M } t_{l \, \pm}^{(1,2) \, \pm} \({\omega \over m} \) 
\nonumber\\
f_{l \, \pm}^{(3) \, \pm} &=&
{m^3 \over F^2 M^2 } t_{l \, \pm}^{(1,3) \, \pm} \({\omega\over m} \)
+{m^3 \over F^4 } t_{l \, \pm}^{(3,3) \, \pm} \( {\omega \over m} \) 
\nonumber\\
f_{l \, \pm}^{(4) \, \pm} &=&
{m^4 \over F^2 M^3 } t_{l \, \pm}^{(1,4) \, \pm} \({\omega\over m} \)
+{m^4 \over F^4 M } t_{l \, \pm}^{(3,4) \, \pm} \( {\omega \over m} \) 
\label{eq:fexpansion2}
\end{eqnarray}
with $m$ the pion mass, $M$ the nucleon mass, $F$ the pion decay constant
and $\omega$ the pion CM energy.
The partial wave unitarity condition 
\be {\rm Im} f_{l \, \pm}^{-1} =
-q,\label{exactunit}
\ee 
where $q$ is the CM momentum, implies that {\em perturbatively} one
has~\footnote{We have checked analytically that the amplitudes in \cite{feme4} are
perturbatively unitary if the following misprints are corrected:
$\displaystyle-\frac{1}{12\omega^2} \frac{\partial
J_0}{\partial\omega}(\omega)$ should read
$\displaystyle\frac{1}{12\omega^2} \frac{\partial
J_0}{\partial\omega}(-\omega)$ in their eq.(3.16) and
$6\omega^4\left(-4M_\pi^2+4\omega^2+t\right)$ should read
$-6\omega^4\left(-4M_\pi^2+4\omega^2+t\right)$ in their eq.(3.18) . In
fact, with these two signs corrected, we reproduce the threshold
parameter expressions given in their eqs.(A.1)-(A.8), except for the $\pi^2$ in
the denominator of the fourth term in the r.h.s. of their (A.8) which
should read $\pi^3$ and the $+g_A \bar d_{18} M_\pi^2/( 4\pi F^2(M_\pi+m))$ in $b^+_ {0+}$,
eq.(A.3),
that should have the opposite sign.}
\begin{eqnarray}
{\rm Im}\, t_{l \, \pm}^{(1,1) \, \pm} = 
{\rm Im}\, t_{l \, \pm}^{(1,2) \, \pm} &=& 
{\rm Im}\, t_{l \, \pm}^{(1,3) \, \pm} = 0 \nonumber \\
{\rm Im}\, t_{l \, \pm}^{(3,3) \, \pm} &=& {q\over m}
 \left[t_{l \, \pm}^{(1,1) \, \pm}\right]^2
\nonumber \\
{\rm Im}\, t_{l \, \pm}^{(3,4) \, \pm} &=& 2{q\over m}
 t_{l \, \pm}^{(1,1) \, \pm}t_{l \, \pm}^{(1,2) \, \pm}
\label{pertunit}
\end{eqnarray}
Following the same ideas as in \cite{GNPR00}, we will consider the 
unitarized amplitude to fourth order:
\bea 
{1\over f |_{\rm Unitarized}} &=&
{F^2 \over m } {1\over t^{(1,1)} + {m\over M} \,
t^{(1,2)}  + \, ({m\over M})^2 \, t^{(1,3)} 
+ \, ({m\over M})^3 \, t^{(1,4)} } 
 -m {{t^{(3,3)} +{m\over M}t^{(3,4)}} \over
[ t^{(1,1)} ]^2 +2 {m\over M}  t^{(1,1)}  
t^{(1,2)}}\label{eq:erajniam} 
\eea 
which, using (\ref{pertunit}) yields immediately (\ref{exactunit}). 
Let us recall that our Improved IAM formula at third order 
read \cite{GNPR00}:
\bea
{1\over f |_{\rm Unitarized}} &=&
{F^2 \over m } {1\over t^{(1,1)} + {m\over M} \,
t^{(1,2)}  + \, ({m\over M})^2 \, t^{(1,3)} 
}  -m {{t^{(3,3)}} \over
[ t^{(1,1)} ]^2 }\label{eq:erajniam3rd} 
\eea
which can be now reobtained from eq.(\ref{eq:erajniam}) by removing the $t^{(1,4)}$ and 
$t^{(3,4)}$ terms  and, consistently with unitarity, 
removing also the $2(m/M) t^{(1,1)}t^{(1,2)} $ factor
 in the second denominator. 
Hence, as we  have 
stressed in the introduction, the knowledge of $t^{(1,4)}$ and 
$t^{(3,4)}$ allows us to test our power counting by including one more 
 term both in the $\od(F^{-2})$ and  $\od(F^{-4})$ contributions.

\section{The third order and the LEC power counting}

In the literature there are two $O(q^3)$ calculations \cite{Mo98,FMS98},
using different choices of counterterms and renormalization schemes,
but only one at $O(q^4)$~\cite{feme4} following the \cite{FMS98} scheme. 
The translation between them does
not simply amount to a change of notation, but involves
some $1/M$ corrections.
Since our results at third order~\cite{GNPR00} were constructed directly
from \cite{Mo98}, we have to check to
what extend our previous $\od(q^3)$ results are reproduced when
using the amplitudes and notation of Refs.~\cite{FMS98,feme4}. In
 so doing two remarks are in order:

First, already at third order, the re-expanded amplitudes of
\cite{Mo98} and \cite{FMS98} differ slightly due both to a different
choice of the reference frame and of the nucleon wave function
renormalization (see comments in \cite{FMS98}). In practice, this just
means that there are slight numerical differences between the
perturbative results, which eventually could be absorbed in the
numerical values of the LEC of the HBChPT Lagrangian.

\begin{figure}
\begin{center}                                                                
\leavevmode
\centerline{\epsfig{file=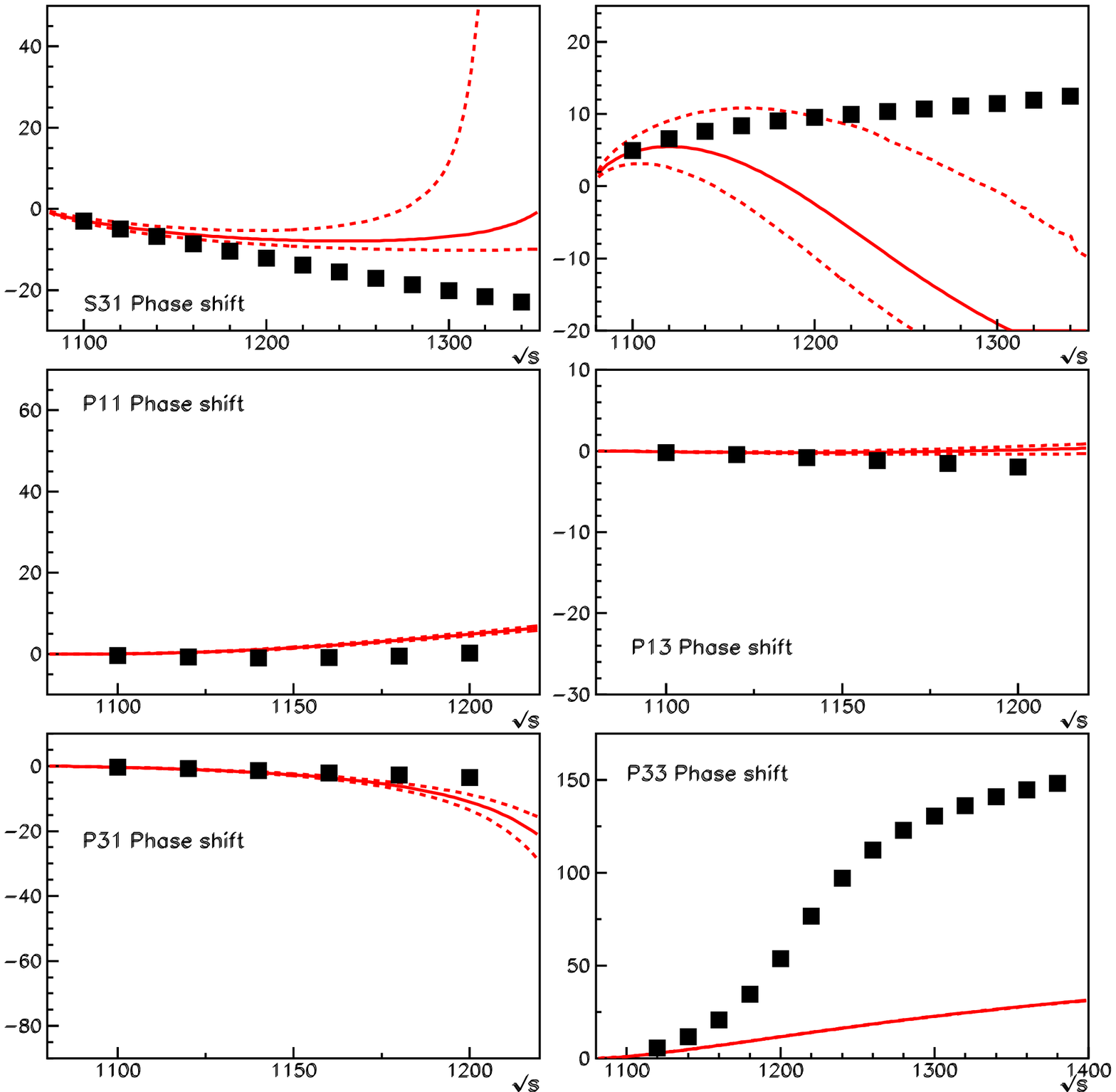,width=9cm},
\epsfig{file=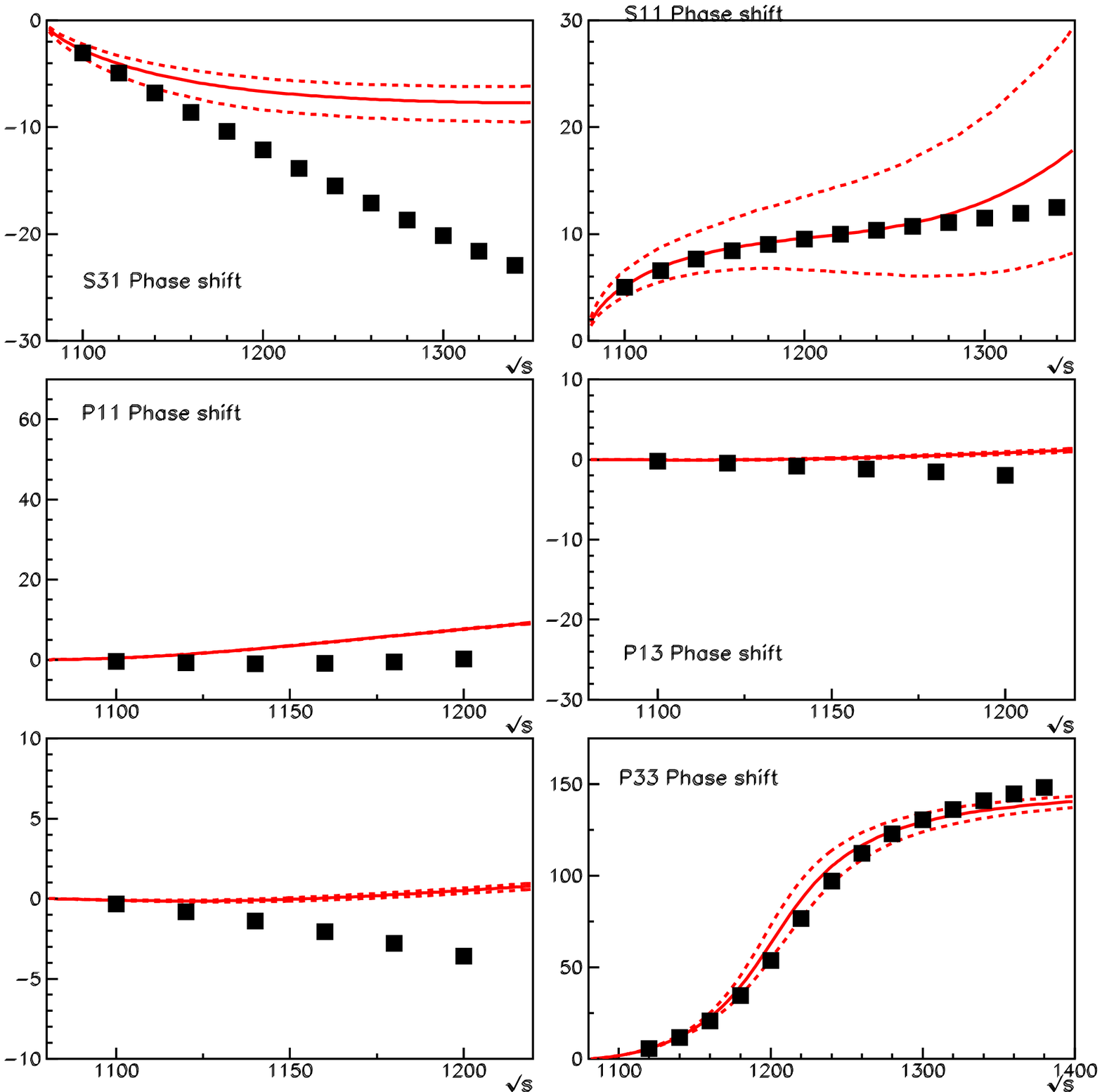,width=9cm}}
\end{center}
\caption[] {\label{ho3} \footnotesize $\od(q^3)$ Unitarized 
phase shifts as a function of
the total CM energy $\protect\sqrt s$: As explained in section III: a) In the two left columns we
use  the $\bd_i$, whereas in  b) in the two right columns we use 
 the $b_i$ set. Experimental data are from
\cite{AS95}. 
The areas between dotted lines correspond to the propagated errors of the 
parameters of fit 1 in ~\cite{FMS98} in both cases.}
\end{figure}

A second point becomes more relevant for our purposes: if we just take
 the third order amplitudes in \cite{FMS98}, we re-expand them
 separating the different contributions and use our third order
 unitarized formula (eq.(13) in \cite{GNPR00}), we find a much worse
 result than in \cite{GNPR00}, particularly in the P33 channel where
 the $\Delta (1232)$ resonance should appear. This is shown in
 Figure \ref{ho3}a.  We remark that we
 are using a set of parameters compatible with those in \cite{GNPR00},
 where the description of the resonance was excellent within the
 errors even without fitting.

The origin of this apparent discrepancy is that we have not taken into
 account that in our power counting scheme, the LEC themselves may
 have contributions of different orders.  In fact, all the difference
 with \cite{GNPR00} is that we have chosen now a different {\it
 parametrization} of LEC, although the {\it numerical} values are
 compatible: in \cite{GNPR00} we followed \cite{EM96,Mo98} where the
 five $\od(q^3)$ LEC appearing in the amplitude are called
 $b_1+b_2,b_3,b_6,b_{16}-b_{15}, b_{19}$.  Here, we follow
 \cite{FMS98,feme4} where the relevant $\od(q^3)$ constants are
 $\bd_1+\bd_2,\bd_3,\bd_5, \bd_{14}-\bd_{15}, \bd_{18}$. Comparing the
 Lagrangian given in equations (2.45)-(2.47) of \cite{FMS98} with that
 in \cite{EM96} one observes that the $b_i$ are related to the $\bd_i$
 for $i=1,2,3$ and to the $\bd_{i-1}$ for $i=6,15,16,19$ typically as:
\be
M^2\bd_i \sim constant + b_i M^2/(16\pi^2 F^2)
\label{reordering}
\ee
with a constant that
 is $\od(1)$ in the $F^{-2}$ counting.  This comes from the fact that
 in \cite{EM96} some finite terms coming from renormalization have
 been absorbed in the $b_i$.

Now, following our power counting arguments, if we replace in the
amplitudes of refs.\cite{FMS98,feme4} the $\bar d_i$ using eq.(\ref{reordering}), there are
pieces in $t^{(1,3)}$ shifted to $t^{(3,3)}$ (remember that all the
dependence with the $\bd_i$ is in $t^{(1,3)}$). This changes the
functional dependence of $t^{(3,3)}$, including, for instance, higher order
polynomial contributions that otherwise were not present. Although
the perturbative amplitude remains the same, the unitarized one changes
since  $t^{(1,3)}$ and $t^{(3,3)}$ are treated on
a different footing. With this procedure we obtain the unitarized results shown in
Figure \ref{ho3}b. The improvement is clear for the P33 wave and
the results are similar to those in \cite{GNPR00}. The corresponding
values for the mass and width of the $\Delta (1232)$ extracted from
the phase shifts are given in the second column of Table
\ref{tab:deltapa}. This highlights the importance of taking into
account the counting of the LEC.

The above separation is, of course, arbitrary, since nothing
prevents us from normalizing the $\bd_i$, which are quantities of
dimension $E^{-2}$ as $(4\pi F^2)\bd_i$ instead of $M^2\bd_i$,
assuming that both are quantities of natural size. Thus, the most
general way to proceed would be to consider as free parameters the
coefficients of the $\od(1)$ and $\od(F^{-2})$ terms in $M^2\bd_i$.
In such a way we would duplicate the number of $\od(q^3)$
LEC, rendering the approach unnecessarily complicated, since we already
know that it is enough to consider
the separation given by the $b_i$ parametrization \cite{EM96}.
We will thus use only that separation in our calculations, but,  after
performing the fit, we will give the results in the $\bd_i$ set for
easier comparison with the literature.

\begin{figure}
\begin{center}                                                                
\leavevmode
\centerline{\epsfig{file=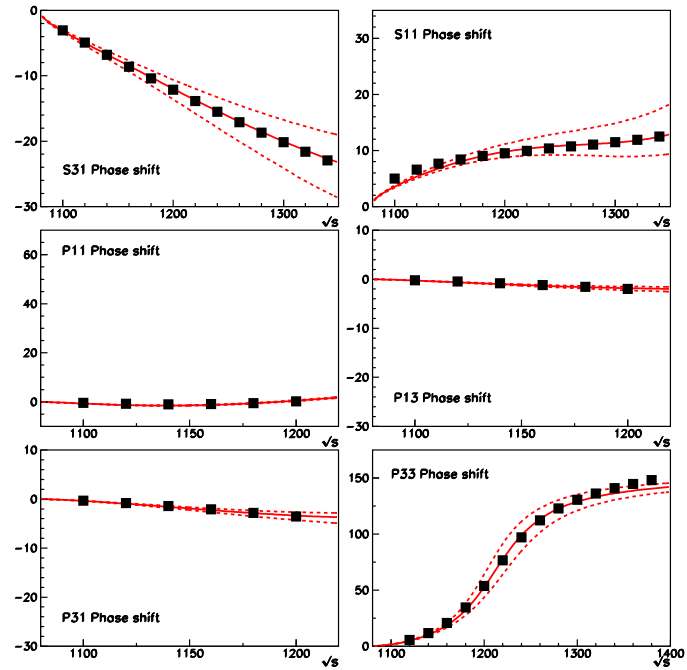,width=9cm}}
\end{center}
\caption[] {\label{fit3} \footnotesize Unitarized $\od(q^3)$ fit. The 
 fit parameters are given in Table \ref{tab:lec2}.
The areas between dotted lines correspond to the errors given in Table 
 \ref{tab:lec2}.}
\end{figure}

 The results of our  $\od(q^3)$  fit is shown in 
 Figure \ref{fit3}. The fit parameters and their errors are given 
 in the fourth column of Table \ref{tab:lec2} whereas 
the results for the $\Delta$ mass and width are given 
 in Table \ref{tab:deltapa}. All them are in 
 agreement with what we found in \cite{GNPR00}. The description of data in 
 all channels is very good and our $\od(q^3)$ LEC are all 
 of natural size although, as it also happened in \cite{GNPR00}, 
 they differ somewhat from those obtained from HBChPT (second column in 
 Table \ref{tab:lec2}). Note that systematic errors are not given in Table III,
although they are dominant  as it can be seen from Table II). 
This is probably another consequence of the poor 
 convergence of the HBChPT series. Let us nevertheless recall that, in 
 \cite{GNPR00} it was shown that one can perform $\od(q^3)$ fits where the 
$\od(q^2)$ parameters $c_i$ are 
 fixed to the predictions of Resonance Saturation \cite{BKM97} and the 
 results are still in excellent agreement with data.

\section{Fourth order results}

Let us consider now  our fourth order unitarized amplitude 
 (\ref{eq:erajniam}) with the
 HBChPT amplitudes of \cite{feme4}.  
In principle, the $\od(q^4)$ amplitude depends on nine different 
combinations of $\od(q^4)$ 
constants $\eb_i$, in addition to the four $\od(q^2)$ $c_i$ and the 
 five  $\od(q^3)$ $\bd_i$. These nine combinations are displayed in 
 the first column of Table \ref{tab:lec2}. 
 However, as noted in \cite{feme4}, the last four combinations actually 
 amount to renormalize the $c_i$, giving rise to new  
$\tilde c_i$ as given in eq.(3.23) of \cite{feme4}. Strictly speaking, the 
 $\od(q^4)$  amplitude depends on 18 parameters, since the $c_i$ still
 appear in the pure $\od(q^4)$ terms. However, replacing those $c_i$ by 
 $\tilde c_i$  introduces higher order corrections, so that the 
 number of free parameters up to $\od(q^4)$ is really 14. At this point, as
 commented in \cite{feme4} one can follow two different  strategies. 
 The first one is to consider as free parameters $\tilde c_i, \bd_i$ and 
 the $\eb_i$ with $i=14-18$. This is the parameter set listed in Table 
 \ref{tab:lec1}. Although this is the more natural set, 
 it has the inconvenience that one cannot disentangle which part of the
 $\tilde c_i$ comes from $\od(q^4)$ renormalization, since those corrections
 are relatively large (another signal of the HBChPT bad convergence). 
As a  consequence,
 it becomes more difficult to compare with previously published values for
 the $c_i$. The alternative (strategy 2) is to {\em fix} the $c_i$ values, 
 which in turn are the ones less subjected to uncertainties and then use
 the $\bd_i$ and the nine combinations of $\eb_i$ as free parameters. This 
second strategy is  useful for instance to fix the 
 $c_i$ to the predictions of Resonance Saturation \cite{BKM97} as we also
 did in \cite{GNPR00}.

In addition, we have to face again the problem of the LEC counting,
 according to the discussion in the previous section. Thus, besides
 the ``reordering'' of terms coming from the $\bd_i$, we now have to
 consider also that coming from the separation of $M^3\eb_i$ into a
 constant (a contribution to $t^{(1,4)}$) plus an $\od\left(M^2/(4\pi
 F)^2\right)$ term (which contributes to $t^{(3,4)}$). Recall that in
 this case we do not have any ``natural'' way to perform that
 separation, as in the $\od(q^3)$ case.

\subsection{The unitarized partial waves to $\od(q^4)$}

\begin{figure}
\begin{center}                                                                
\leavevmode
\centerline{\epsfig{file=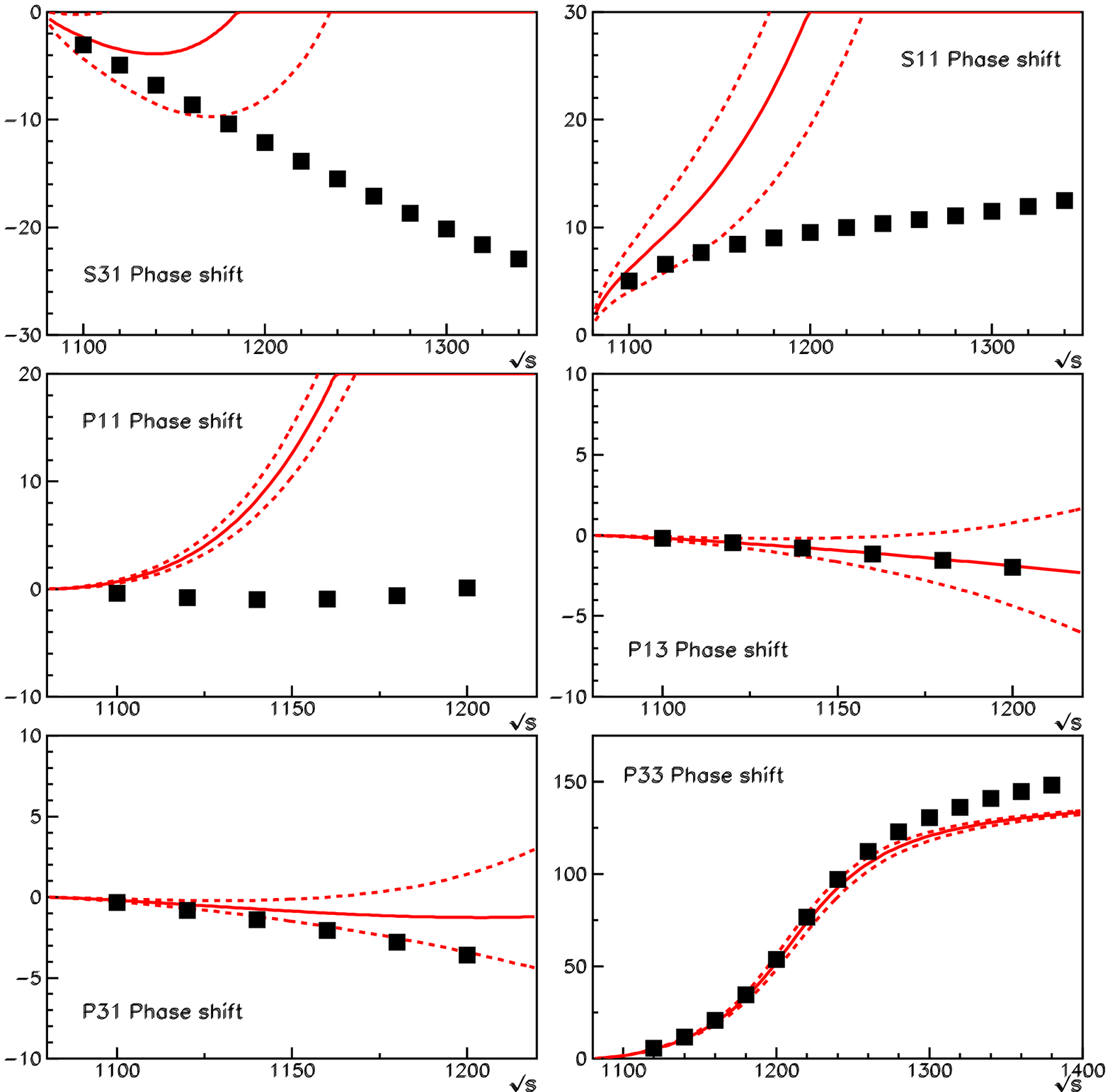,width=9cm},
\epsfig{file=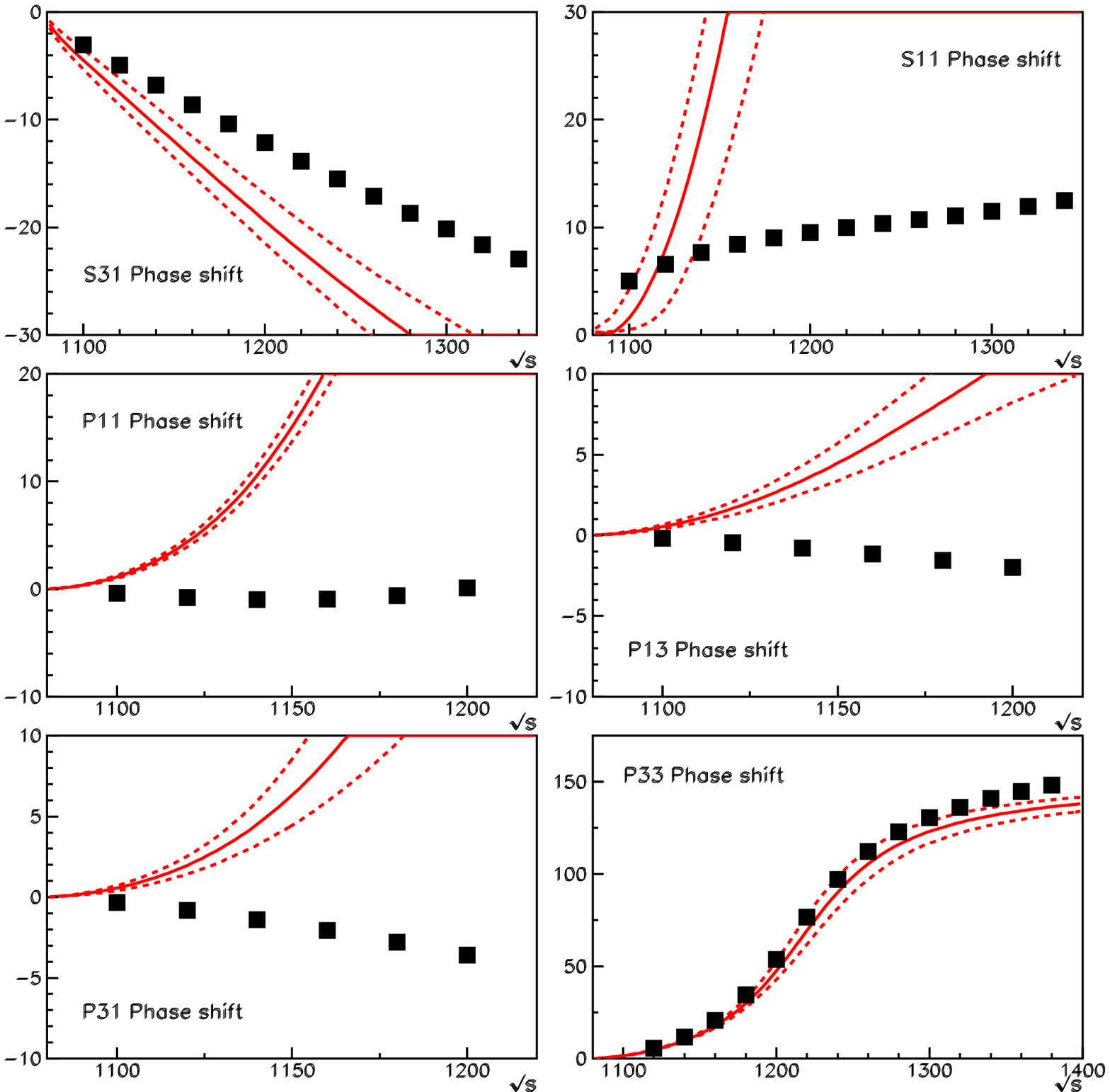,width=9cm}}
\end{center}
\caption[] {\label{ho4} \footnotesize $\od(q^4)$ Unitarized 
phase shifts as a function of
the total CM energy $\protect\sqrt s$. a) The two left columns 
correspond to reordering the $\bd_i$ only, whereas 
b) in the two right columns we have  reordered both $\bd_i$ and
 $\eb_i$.
Experimental data are from \cite{AS95}. 
The areas between dotted lines correspond to the propagated errors of the 
parameters of fit 3 in ~\cite{feme4} with bigger errors (see Sect.IV.a)}
\end{figure}

First, as we did to $\od(q^3)$, we will show the predictions of our
 formula {\em without} fitting, performing a Monte Carlo sampling of
 the perturbative LEC, assuming that they are uncorrelated.  Following
 the first strategy, we have used the LEC given in \cite{feme4}, in particular
 those given in their ``Fit 3'' that we reproduce in Table \ref{tab:lec1}. 
We also list the  ``Fit 1 and 2'' parameter sets to illustrate that, as pointed out in \cite{feme4},
the systematic errors are much larger than the statistical ones, 
that we are quoting in the Table. 
 For that reason we will
 take bigger errors, since the errors listed in \cite{feme4} are
 clearly underestimated.  In view of the uncertainties in \cite{feme4}
 we have assigned an error of $1.0$ to
 $\eb_{14},\eb_{15},\eb_{17},\eb_{18}$, of $0.5$ to $\tilde c_1$,
 $\tilde c_2$, $\tilde c_3$ (the $\tilde c_i$ have bigger
 uncertainties than the $c_i$ due to their $\eb_i$ contribution) and
 of $0.25$ to the remaining LEC. In Figure \ref{ho4}a we show the
 $\od(q^4)$ prediction ``redefining'' the $\bd_i$ as before but
 without doing so for the $\eb_i$, while in Figure \ref{ho4}b we
 have also redefined the $\eb_i$ for convenience as $M^3 \eb_i = 1+
 \ef_i M^2/(4\pi F)^2$. Throughout this paper, and for practical
 purposes, we will consider only these two situations.

In view of Figure \ref{ho4} there are several comments in order:
 First, consider the P33 channel, where our approach is meant to be more
 accurate. Here our $\od(q^4)$ result confirms the $\od(q^3)$ one and even
 improves it slightly. Observe for instance the results for the $\Delta$ 
 parameters given in the fourth column of Table \ref{tab:deltapa}, 
corresponding to Figure \ref{ho4}b. This is one of the main conclusions
 of this work, namely that the $\od(q^4)$ calculation confirms that our
 unitarization method generates dynamically the $\Delta (1232)$ resonance. 
The improvement of the P33 channel description
 is also a common feature with \cite{feme4}. Note also that this conclusion
 does not change by using the $\eb_i$ or the $\ef_i$ formulas as long as 
 we redefine the $\od(q^3)$ LEC as before.

As for the other channels, we see no actual improvement when comparing 
 with the unfitted $\od(q^3)$ in 
 Figure \ref{ho3}b. On the contrary, we get worse results for most of them,
 especially the two $S$ channels and the $P_{11}$ one. 
 Here, we see significant differences between using the $\ef_i$ prescription
 or not. In fact, without fitting, 
our choice for the $\ef_i$ does not seem to give better
  results than using the  $\eb_i$ directly
(Figure \ref{ho4}a) or, equivalently, neglecting the $\od(F^{-2})$
 contribution in  $M^3 \eb_i$. 

The hope is that we can perform $\od(q^4)$ fits which 
 improve these five channels without spoiling the P33 one and with a 
 reasonable size for the LEC. However, we must bear in mind that, 
as commented before, 
the low-energy $\od(q^4)$  fits performed in \cite{feme4} already show that
 one gets only slightly  better descriptions  and bigger
 uncertainties for the LEC than the $\od(q^3)$.

\begin{table}
  \begin{center}
    \begin{tabular}{|c|c|c|c|c|c|}
&$\od(q^3)$ unfitted& fit $\od(q^3)$  & $\od(q^4)$ unfitted & 
fit $\od(q^4)$ & PDG\\\hline\hline
\rule[-.3cm]{0cm}{.8cm}
$M_\Delta $  (MeV)& 1221 $\er{11}{10}$ 
&1229  $\er{14}{12}$&1238$\er{10}{9}$       
& 1232$\er{35}{29}$   & 1230 - 1234\\ \hline
\rule[-.3cm]{0cm}{.8cm}
$\Gamma_\Delta$ (MeV)&  111.2$\er{16.9}{14.3}$ 
& 108.4 $\er{20.6}{16.5}$&125.2$\er{20.4}{16.4}$& 107.3 $\er{45.1}{31.0}$ 
&  115-125 
    \end{tabular}
  \end{center}
\caption{$\Delta (1232)$ resonance parameters in the different cases 
considered in this paper. The resonance mass and width are
obtained from the condition $\delta^1_{33}|_{s=M_\Delta^2}=\pi/2$ and
$1/\Gamma_\Delta=M_\Delta (d\delta^1_{33}/ds)\vert_{s=M^2_\Delta}$.}
\label{tab:deltapa}
\end{table}

\subsection{$\od(q^4)$ fits}

\begin{figure}
\begin{center}                                                                
\leavevmode
\centerline{\epsfig{file=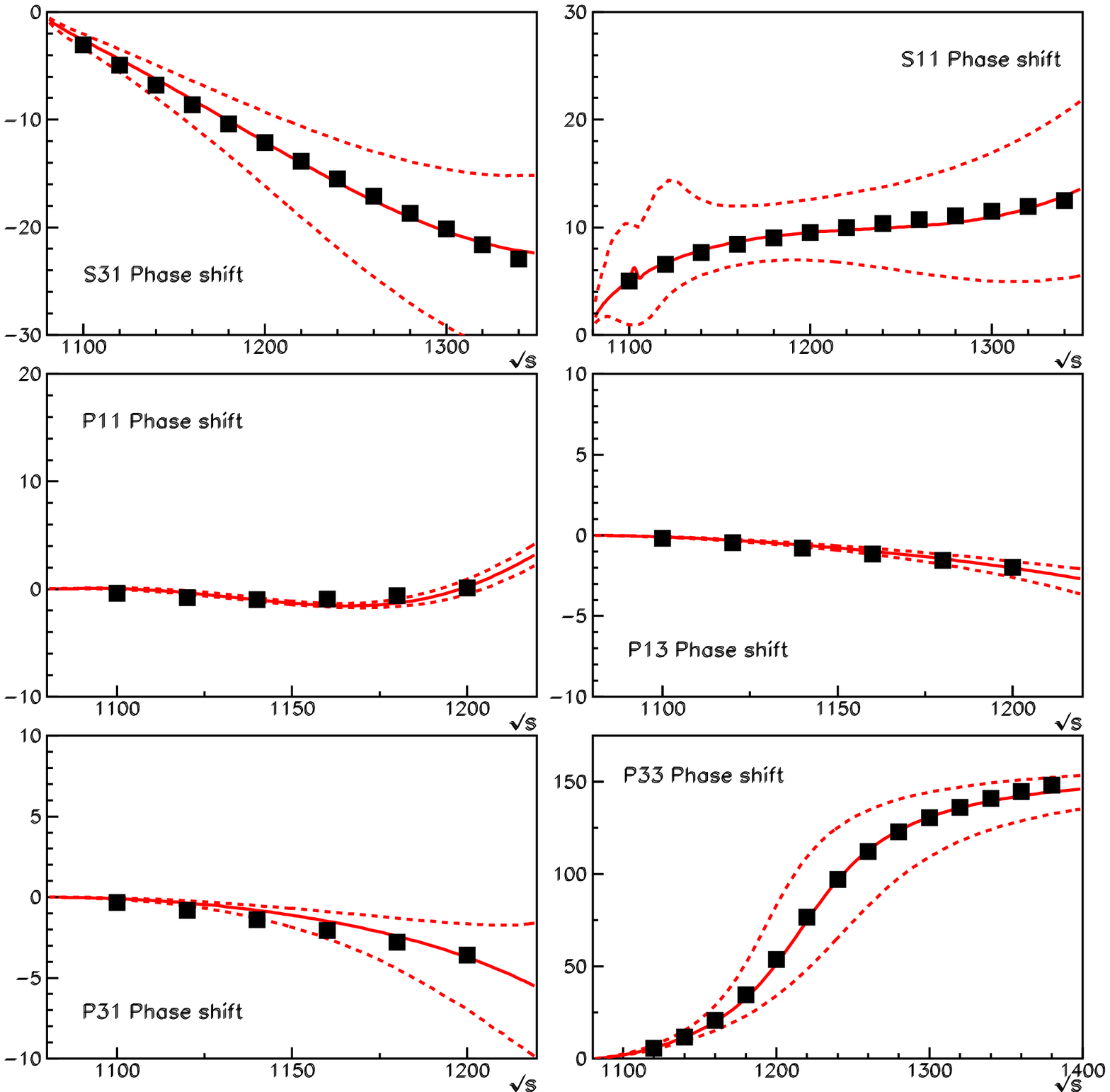,width=9cm},
\epsfig{file=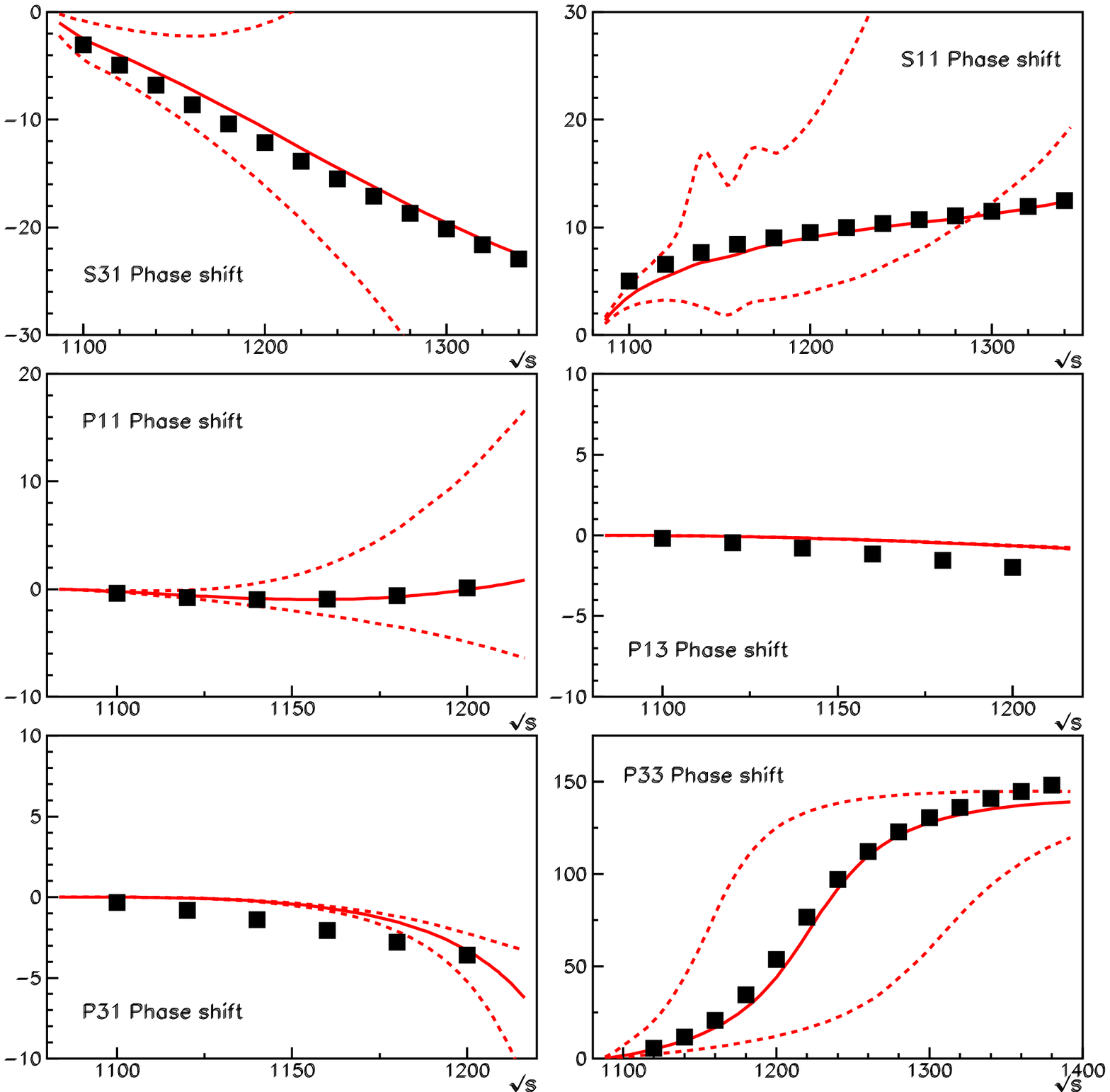,width=9cm}}
\end{center}
\caption[] {\label{fit4} \footnotesize Unitarized $\od(q^4)$ fits:
a) using strategy 1 in the two left columns and b) using strategy 2
in the two right columns.  Experimental data are from \cite{AS95}. The 
 fit parameters and errors  are given in Tables \ref{tab:lec1} and 
 \ref{tab:lec2} respectively. 
The areas between dotted lines correspond to propagating those errors.}
\end{figure}

In Figure  \ref{fit4}a we show the result of our best fit with the fit errors 
propagated. Here we have 
 followed the first strategy and we have used the $\ef_i$ defined in the 
 previous section. The LEC and their errors are given in 
the last column in Table \ref{tab:lec1}. The main observation is that we 
 reproduce the data with constants of natural size. The constant 
 $\eb_{18}$ turns out to be highly 
correlated numerically with $\eb_{17}$ so that 
 we have chosen to fix one of them to the perturbative value. 
For the $\Delta$ parameters
 we get the results in the fifth column in Table \ref{tab:deltapa} which is 
 still fully compatible with the experimental result. Note that, as expected
 from our previous comments, the uncertainties in the fit parameters are
 now bigger than in the $\od(q^3)$ fit. However, the quality of the fit is
 comparable if not better: we get a $\chi^2/d.o.f \sim 0.3$ 
for the $\od(q^3)$ fit in Figure \ref{fit3} and 
 $\chi^2/d.o.f \sim 0.17$ for that in Figure \ref{fit4}a. 
As is customarily done \cite{FMS98,GNPR00}, for the $\chi^2$ calculation 
we havee added some error to the data; in particular, we have
chosen to add a 3\%  relative error plus one degree systematic error.
Therefore, our 
 method shows clear signs of convergence when we perform unconstrained fits, 
 although the uncertainties in the LEC remind us of the bad convergence of 
 the HBChPT series. Note also that the bigger uncertainties are in the $S$
 channels, as we have commented before.

We also show, in Figure \ref{fit4}b,  the result of a fit using 
 the second strategy and fixing the  $c_i$ 
as the central values of the predictions of Resonance Saturation 
 \cite{BKM97}. As it also happened with the $\od(q^3)$ in \cite{GNPR00}, 
 the fit result is slightly worse when the $c_i$ are not free parameters. 
 Here we obtain a better fit when using directly the $\eb_i$ and not
 the $\ef_i$ as free $\od(q^4)$ parameters. 
 Nevertheless, some  of the constants become of 
unnatural size and also the errors are bigger than for the unconstrained fit. 
For the fit in Figure  \ref{fit4}b we get a 
$\chi^2/d.o.f \sim 0.58$ and the LEC  listed in the fifth column in 
 Table \ref{tab:lec2}. The correlations between the different LEC 
also become more important. Here, in addition to 
$\eb_{17}$ and $\eb_{18}$, there are also  strong 
 correlations among  $\eb_{14}$,  $\eb_{15}$, $\eb_{16}$, 
$ \eb_{22}-4\eb_{38}$,   $\eb_{20}+\eb_{35}-g_A \bd_{16}/(8M)$, 
 $ 2\eb_{21}-\eb_{37}$    and 
 $ 2\eb_{19}-\eb_{22}-\eb_{36}$ which allow to fix one of them. 
It should be commented that one could find fits with more natural
 values and a higher $\chi^2/d.o.f$ but we have preferred to show the best fit, 
 emphasizing the convergence problems.

\begin{table}
\begin{center}
\begin{tabular}{|c||ccc||c|}
& \multicolumn{3}{c||}{Fettes-Mei{\ss}ner  \protect{\cite{feme4}}}&
Our IAM\\  
& Fit 1& Fit 2& Fit 3& $\od(q^4)$ fit\\\hline
$ \tilde c_1$ 
&$-2.54\pm0.03$
&$-0.27\pm001$
& $-$3.31$\pm$ 0.03  
& $-$ 1.43 $\pm$ 0.10  
\\
$\tilde c_2 $ 
&$0.60\pm0.04$
& $3.29\pm0.03$
& 0.13 $\pm$ 0.03     
& $-$ 0.33 $\pm$ 0.10
\\
$\tilde c_3 $  
&$-8.86\pm0.06$
&$-1.44\pm0.03$
& $-$10.37 $\pm$ 0.05  
& $-$ 2.62 $\pm$ 0.17
 \\
$\tilde c_4 $  
&$2.80\pm0.13$
&$3.53\pm0.08$
& 2.86$\pm$0.10       
& 0.64$\pm$0.10  
\\
$\bd_1 + \bd_2 $    
&$5.68\pm0.09$
&$4.45\pm0.05$
& 5.59$\pm$ 0.06  
& 1.11 $\pm$ 1.02  
\\
$\bd_3 $ 
&$-4.82\pm0.09$
&$-2.96\pm0.05$           
&$-$4.91$\pm$0.07  
& $-$0.59$\pm$1.06
   \\
$\bd_5 $   
&$-0.09\pm0.06$
&  $-0.95\pm0.03$            
& $-$0.15$\pm$0.05    
& $-$0.19 $\pm$ 0.40 
  \\
$\bd_{14}-\bd_{15}$ 
&$-10.49\pm0.18$
&    $-7.02\pm0.11$
& $-$11.14 $\pm$ 0.11    
&  2.49 $\pm$1.93
     \\
$ \bd_{18}$ 
&$-1.53\pm0.17$
& $-0.97\pm0.11$                   
& $-$0.85$\pm$0.06 
& $-$ 17.49$\pm$ 1.31
 \\
$ \eb_{14}$  
&$6.39\pm0.27$
&$-4.68\pm0.14$                   
& 7.83$\pm$0.23 
& 1.58 $\pm$ 0.53
\\
$ \eb_{15}$  
&$4.65\pm0.31$
&$-18.41\pm0.15$                     
& 9.72$\pm$0.25 
& $-$ 1.41 $\pm$ 1.13
 \\
$ \eb_{16}$    
&$7.05\pm0.30$
&$7.79\pm0.15$     
& 6.42$\pm$ 0.25 
& 3.50 $\pm$ 1.30
 \\
$ \eb_{17}$    
&$4.88\pm0.98$
&$-17.79\pm0.53$                   
& 5.47$\pm$ 0.64 
& 6.56 $\pm$ 1.92
 \\
$ \eb_{18}$  
&$-9.15\pm0.98$
&$19.66\pm0.53$                   
& $-$0.17$\pm$0.64 
& $-$0.17 (fixed)
 \\
\end{tabular}
\end{center}
\caption[pepe]{\footnotesize HBChPT low energy constants with strategy 1.
 The
 $c_i$, $\bd_i$ and $\eb_i$ are given in GeV$^{-1}$,GeV$^{-2}$
 and GeV$^{-3}$ respectively. We list the three parameter sets provided 
in \cite{feme4}
to illustrate the large systematic uncertainties already existing in the perturbative determinations.}
\label{tab:lec1}
\end{table}

\begin{table}
\begin{center}
\begin{tabular}{|c||c|c||c|c|}
&  Fettes \emph{et al.} \protect{\cite{FMS98}}
&Fettes-Mei{\ss}ner  \protect{\cite{feme4}} 
& $\od(q^3)$ fit&$\od(q^4)$ fit\\
& & 
 $\od(q^4)$ (Strategy 2) & &RS $c_i$ of 
 \cite{BKM97}
 \\ \hline 
$ c_1$  
& $-$1.53$\pm$  0.18  
& $-$1.47 (input)  
& $-$0.43 $\pm$  0.04
& $-$0.9 (input)
\\
$c_2 $  
& 3.22 $\pm$ 0.25     
& 3.26 (input) 
& 1.28$\pm$0.03 
&3.9 (input)
\\
$c_3 $  
& $-$6.20$\pm$ 0.09  
& $-$6.14 (input) 
& $-$3.10 $\pm$ 0.05
& $-$5.3 (input)
 \\
$c_4 $  
& 3.51$\pm$0.04       
&  3.50 (input) 
&  $1.51\pm$0.04
&3.7 (input)
\\
$\bd_1 + \bd_2 $    
& $2.68\pm 0.15$   
& $4.90 \pm 0.05$ 
& $2.66\pm 0.20$
& 10.36 $\pm$0.53
\\
$\bd_3 $                 
&$-$3.11$\pm$0.79  
& $-$4.19$\pm$0.07
& $-0.32\pm 0.2$
& $-$ 4.07 $\pm$0.26
   \\
$\bd_5 $                 
& 0.43$\pm$0.49    
& $-0.16\pm 0.05$ 
& $-1.66 \pm 0.10$
& $-$ 3.23 $\pm$ 0.31
  \\
$\bd_{14}-\bd_{15}$     
& $-5.74 \pm 0.29$    
& $-9.31 \pm 0.10$ 
& $-5.34 \pm 0.40$
& $-$1.17 $\pm$ 0.68
     \\
$ \bd_{18}$                     
& $-$0.83$\pm$0.06 
& $-$0.84$\pm$0.06 
& $-2.60 \pm$0.20
& $-$ 53.29 $\pm$ 3.37
 \\
$ \eb_{14}$                     
&  
& $4.19\pm0.23$ 
& 
& 2.24 $\pm$ 0.94
\\
$ \eb_{15}$                     
&  
& $4.54\pm0.25$ 
& 
& $-$ 2.17 $\pm$ 2.14
 \\
$ \eb_{16}$                     
&  
& $2.74\pm 0.24$ 
& 
& 5.12 $\pm$ 1.26
 \\
$ \eb_{17}$                     
&  
& $7.20\pm 0.64$ 
& 
&$-$ 0.95 $\pm$ 1.86
 \\
$ \eb_{18}$                     
&  
& $-$ 3.36$\pm$0.64 
& 
& $-$ 3.36 (fixed)
 \\
$ \eb_{22}-4\eb_{38}$                     
&  
& 27.72$\pm$0.74 
& 
& 27.72 (fixed)
 \\
$\eb_{20}+\eb_{35}-g_A \bd_{16}/(8M)$                     
&  
& $-$17.35$\pm$0.36 
& 
& $-$ 70.11 $\pm$ 3.07
 \\
$ 2\eb_{19}-\eb_{22}-\eb_{36}$                     
&  
& $-25.12\pm0.69$ 
& 
&  97.14 $\pm$ 10.00
 \\
$ 2\eb_{21}-\eb_{37}$                     
&  
& $-$5.00$\pm$1.43 
& 
& 17.64 $\pm$10.7
\\
\end{tabular}
\end{center}
\caption[pepe]{\footnotesize HBChPT low energy constants with strategy 2. The
 $\tilde c_i$, $\bd_i$ and $\eb_i$ are given in GeV$^{-1}$,GeV$^{-2}$
 and GeV$^{-3}$ respectively.}
\label{tab:lec2}
\end{table}

\section{Conclusions}

In this work we have used a unitarized $\pi N$ scattering amplitude
including up to $\od(q^4)$ terms in the standard Heavy Baryon Chiral
Perturbation Theory expansion. This has allowed to test our method of
considering the $F^{-2}$ expansion resumming the $M^{-1}$
contributions.  The description of the $P33$ channel and the $\Delta$
resonance, which is dynamically generated, are excellent within the
experimental errors. The inclusion of the new $\od(q^4)$ terms does
not change much this picture, which is a consistency check of our
formalism.

In order to describe accurately the data in the other three $P$
channels and the two $S$ ones, one needs to fit the LEC. We have
showed that it is possible to fit the six channels simultaneously with
natural values for the LEC, although with considerably larger
uncertainties than in the $\od(q^3)$. This is a consequence of the
poor HBChPT convergence which shows up already at the pure
perturbative level at lower energies.  In fact, when one tries to
perform $\od(q^4)$ fits constraining the lowest order constants to the
Resonance Saturation hypothesis, some of the LEC become of unnatural
size and their errors increase considerably.  These convergence
problems are especially important in the two $S$ waves.
 
We have also discussed the issue of the LEC power counting, which is
relevant in our expansion scheme. The importance of this effect has
been especially highlighted in the $P33$ channel, where a correct
splitting of the $\od(q^3)$ constants is crucial. To $\od(q^4)$ the
influence of such LEC reordering is smaller as far as the resonance is
concerned, although it may improve the convergence in the other
channels.

In summary, our unitarization method is robust and is almost not
affected by the HBChPT convergence problems as far as the generation
of the $\Delta(1232)$ resonance is concerned. However, the predictions
of the unitarized amplitude to fourth order for other channels show
similar problems of convergence as the perturbative one, even though
one can still find excellent descriptions of data with natural values for the
low-energy constants. It seems a natural continuation of this work to
implement our unitarization methods within the context of the Lorentz
invariant formalism proposed in \cite{bele} (see also
Ref.~\cite{Gegelia:1999qt}) which has better convergence properties.

\begin{acknowledgments}
This work is supported in part by funds provided by the Spanish DGI
with grants no. BFM2002-03218, BFM2000-1326 and BFM2002-01003;Junta de Andaluc\'{\i}a grant
no. FQM-225 and EURIDICE  with contract number HPRN-CT-2002-00311. 

\end{acknowledgments}

\end{document}